\documentclass[journal]{IEEEtran}
\usepackage{cite}
\usepackage{hyperref}
\usepackage{caption}
\ifCLASSINFOpdf
\else
\fi
\usepackage{amsmath}
\usepackage{marvosym}
\usepackage{amssymb}
\usepackage{textcomp}
\usepackage{algorithmic}

\usepackage{array}

\title{Fusing Global and Local: Transformer-CNN Synergy for Next-Gen Current Estimation}

\author{
Junlang Huang\textsuperscript{1,*},
Hao Chen\textsuperscript{1,*},
Li Luo\textsuperscript{1,\dag},
Yong Cai\textsuperscript{1,\dag},
Lexin Zhang\textsuperscript{1},
Tianhao Ma\textsuperscript{2},
Yitian Zhang\textsuperscript{1},\\
Zhong Guan\textsuperscript{1,\textsuperscript{\textdaggerdbl}}%
\thanks{\textsuperscript{1} Sun Yat-Sen University, School of Microelectronics Science and Technology.}%
\thanks{\textsuperscript{2} Zhuhai Chipoly Technology Ltd. \texttt{matianhao@chipoly.com.cn}.}%
\thanks{\textsuperscript{*} These authors contributed equally.}%
\thanks{\textsuperscript{\dag} Co-second authors.}%
\thanks{\textsuperscript{\textdaggerdbl}Corresponding author: Zhong Guan (\texttt{guanzh23@mail.sysu.edu.cn}).}%
}

\usepackage{graphicx}

\begin{document}

\maketitle

%
%
%


\begin{abstract}
This paper proposes an end-to-end method for fast voltage waveform prediction based on a hybrid CNN–Transformer architecture. Unlike traditional approaches that iteratively solve the complete RC topology, the proposed method innovatively incorporates the transfer function as a high-level input feature, reducing modeling complexity from $\mathcal{O}(n^2)$ to $\mathcal{O}(n)$ and significantly lowering computational overhead.

To enable generalization to arbitrary-order transfer functions, we exploit the physical property that any high-order transfer function can be decomposed into a sum of first-order and repeated-pole components. Accordingly, we design a two-stage sub-model: a base model predicts the response of the dominant first-order term, while a compensation module consisting of a single correction network directly outputs point-wise correction values. These corrections are additively applied to the base response to generate the complete high-order prediction.

The proposed method eliminates the need for iterative computation or manual simplification, achieving both high modeling efficiency and generalization capability. Experimental results show that, compared to HSPICE, our approach achieves a root mean square error (RMSE) as low as 0.0098 and maintains an RMSE of 0.0095 on high-order networks, demonstrating superior performance in waveform modeling and transfer function generalization.
\end{abstract}


%
\IEEEpeerreviewmaketitle

\section{Introduction}
%
%
%
%
\IEEEPARstart{I}{n} recent years, deep learning technologies have made remarkable progress, with Transformer-based architectures demonstrating exceptional performance and significant advantages in fields such as natural language processing (NLP), computer vision, and time-series data modeling.\cite{vaswani2017attention} Transformer models, by effectively capturing complex relationships and long-range dependencies, offer a novel perspective for data-driven modeling. This technological advancement has inspired researchers to explore its potential applications in traditional engineering domains, especially in complex physical modeling and signal prediction.\cite{devlin2019bert}

As integrated circuit (IC) technology nodes continue to shrink, the nonlinear characteristics of standard cells and the parasitic effects of interconnects become increasingly prominent, significantly increasing the complexity of circuit simulation. The RC current response of signal lines, as a key indicator for evaluating timing, power consumption, and electromigration (EM) reliability, plays a decisive role in overall chip performance. \cite{black1969electromigration}\cite{sharma2011vlsi}To accurately analyze RC current responses and their behavior in high-speed and high-density interconnects, existing simulators typically employ traditional modeling methods. However, these methods face a trade-off between accuracy and computational efficiency, making it challenging to meet the increasingly complex demands of IC design.

\begin{figure}[h]
    \centering
    \includegraphics[width=0.8\linewidth]{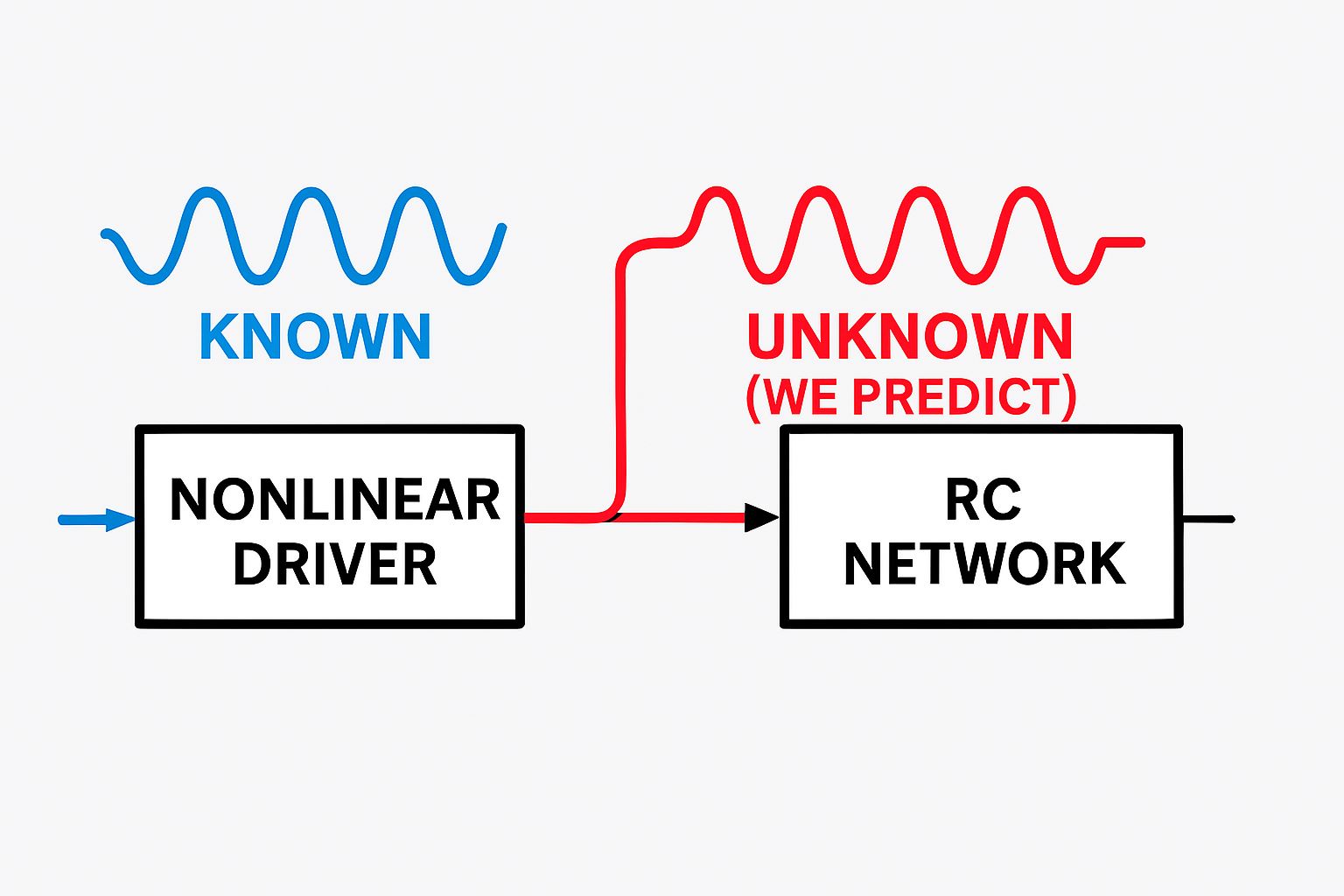}
    \caption{\textbf{Task Description} Given a known input signal, the task is to predict the nonlinear driver's output waveform before it enters the RC network. This intermediate signal, marked in red, is unknown and serves as the prediction target of our model.}
    \label{fig:enter-label}
\end{figure}

Currently, signal line RC response modeling methods can be broadly categorized into three types:

The first category is the Current Source Model (CSM). Criox and Wong proposed a gate cell current source model called Blade \cite{croix2003blade}, which consists of a voltage-controlled current source, internal capacitance, and a one-step time-shift operation. Kellor further enhanced model accuracy by introducing the KTV model \cite{keller2004robust}, which considers Miller capacitance. Subsequently, Li and Acar \cite{li2005waveform} and Fatemi et al. \cite{fatemi2006statistical} introduced input and output parasitic capacitances, modeling the output current source as a function of input/output voltages, gradually incorporating nonlinear characteristics into CSM models. However, since CSM-based methods can only match fixed effective capacitances (up to two) throughout the process, the simulation accuracy of current/voltage waveforms is inherently limited \cite{amin2006multi}\cite{kashyap2007nonlinear}\cite{menezes2008true}\cite{amelifard2008current}\cite{nazarian2010accurate}\cite{katam2019timing} . In recent years, widely adopted industry methods such as Composite Current Source (CCS) \cite{synopsysCCS} and Extended Current Source Model (ECSM) \cite{cadence_ecsm} have established driver and receiver models for each cell to handle scenarios with nonlinear input and crosstalk. Nevertheless, CSM-based approaches still face significant challenges in matching high-order RC load characteristics, limiting their accuracy in current response prediction.

The second category is the Voltage Response Model (VRM), such as the Non-Linear Delay Model (NLDM). Iterative methods \cite{qian1994modeling} \cite{abbaspour2003calculating} \cite{dartu1996performance} \cite{wang2006modeling} , although capable of achieving high precision, often require substantial CPU time for convergence. Non-iterative methods  \cite{kahng1999improved} \cite{shao2003explicit}, on the other hand, rely on closed-form expressions that offer faster computation but can result in output waveform matching errors of up to 15\%  \cite{jiang2010non}. Furthermore, as technology nodes shrink and RC loads become more complex, two-parameter fitting methods struggle to accurately capture the response curve of RC networks, limiting their applicability in high-precision simulations  \cite{garyfallou2021gate}.

The third category consists of Direct Waveform Prediction Methods, such as double exponential functions  \cite{jain2011accurate}, Weibull functions \cite{amin2005weibull}, and gamma functions \cite{lin1998h}, which directly fit the current or voltage response. Recently, a macromodeling method \cite{mirzaie2020macromodeling} was proposed that uses SPICE to extract parameters for modeling, which improves accuracy to some extent. However, these direct fitting methods are unable to predict initial overshoot/undershoot effects, which become more pronounced when the input slope is large.

Against this background, Transformer-based modeling methods show great potential.\cite{vaswani2017attention} Transformer models excel at capturing complex features and nonlinear dependencies, making them well-suited for addressing the shortcomings of traditional methods in complex signal response prediction. By incorporating Transformer models, it is possible to efficiently model signal line RC responses and leverage data-driven approaches to capture dynamic behaviors and detailed characteristics. This study aims to explore the application of Transformers in RC current response prediction, with the goal of improving simulation accuracy and computational efficiency while providing novel solutions for EM analysis and signal integrity evaluation.

 The dynamic behavior of practical circuits is governed primarily by a small number of “dominant poles.” Typically, researchers select the most significant dozen to several dozen poles—based on energy contribution or frequency‐domain gain thresholds—from an original RC network of thousands or even tens of thousands of orders, and then apply Model Order Reduction (MOR) techniques to reduce the network to the corresponding order \cite{odabasioglu2003prima,cheng2012model}. For example, for a 5000th-order RC network, extracting the top 20–50 energy‐contributing poles can reduce the model order to just a few tens of poles, while in most practical cases the frequency response error remains within a few percentage points, thereby greatly lowering model complexity.

Our method seamlessly integrates with the MOR reduction pipeline. On low-order models (e.g., 5th- or 15th-order), the CNN–Transformer architecture achieves a prediction RMSE as low as 0.0098; when applied directly to higher-order reduced models, it still maintains an RMSE of at most 0.0095. Crucially, by exploiting the mathematical fact that any high-order transfer function can be decomposed into a sum of first-order terms, our framework predicts each term’s response independently and then sums them—irrespective of the RC network’s topology—thereby guaranteeing both the universality and consistency of the prediction.

This paper proposes a hybrid architecture combining Transformer and CNN for predicting voltage waveforms in signal lines. Unlike traditional methods such as current source models, driver linear representations, waveform function fitting, or equivalent load capacitance methods, our model does not rely on fixed simplified models of standard-cell drivers or RC loads. Instead, we innovatively introduce the transfer function of the signal line output node as a high-level feature in the modeling process, leveraging the global feature modeling capability of Transformers and the local feature extraction advantages of CNNs to accurately predict sequence responses. By exploiting the decomposability of high-order transfer functions, we construct a generalizable prediction framework that requires training only a small number of sub-models to adapt to high-order RC networks, thus avoiding the cumbersome iterative process and manual modeling involved in traditional methods.

The model adopts a two-stage architecture: the base model predicts the response of the dominant first-order component, while the compensation module further refines the output through a unified correction network. This correction network directly produces point-wise adjustments that are additively applied to the base response, resulting in a complete high-order waveform prediction. Experimental results demonstrate that the proposed method significantly enhances the accuracy of current waveform prediction while effectively reducing modeling complexity and computational overhead.

\begin{figure*}[ht]
    \centering
    \includegraphics[width=\textwidth]{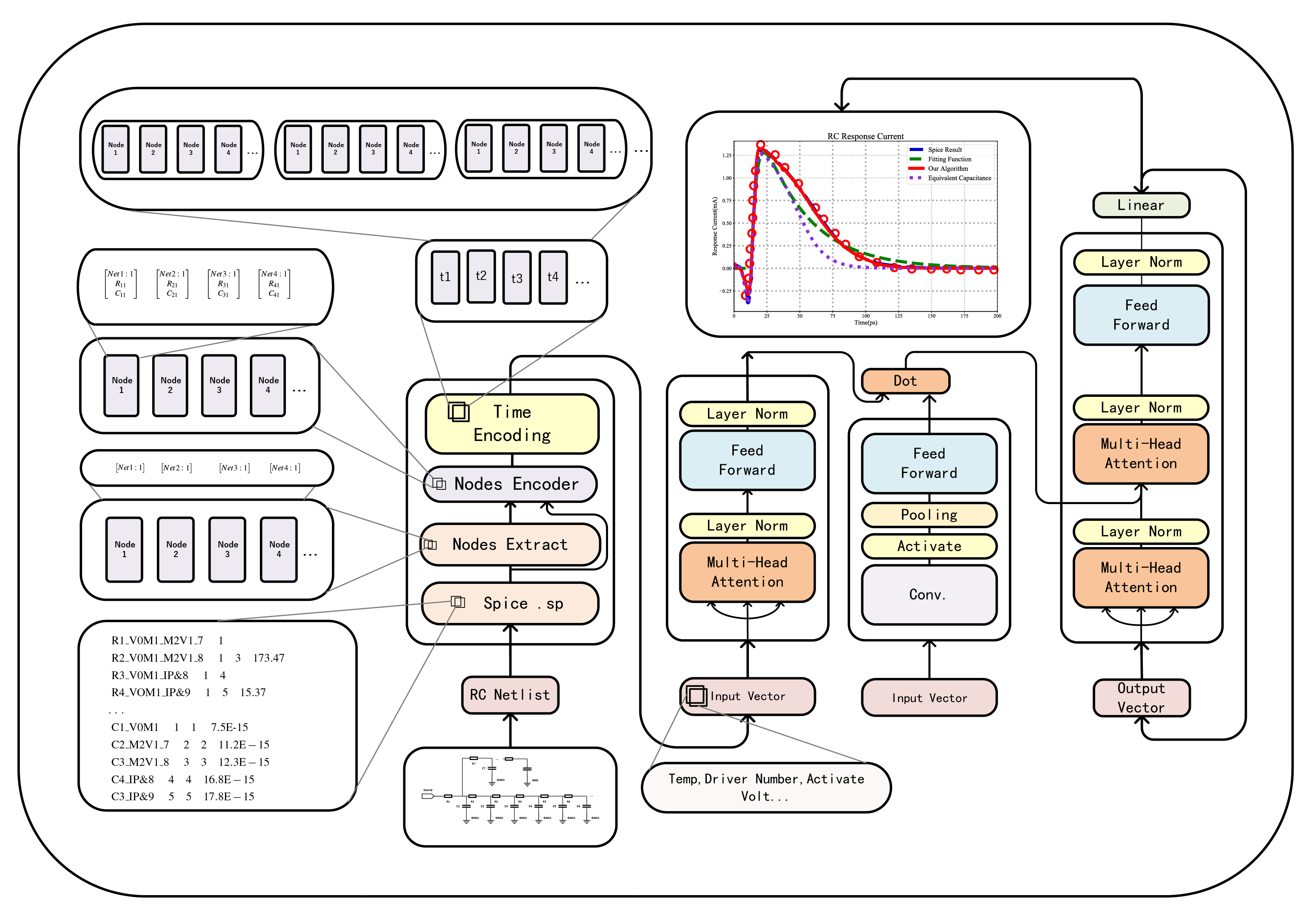}
    \caption{\textbf{Model overview.} We extract structural features from the RC netlist using a node encoder and combine them with time encodings to form the input sequence. This input is processed in parallel by a Transformer branch for global attention and a CNN branch for local feature extraction. The outputs are fused and further refined through Transformer layers to predict the transient current response.}
    \label{fig:model-overview}
\end{figure*}

The organization of this paper is as follows:  
Section II briefly introduces the theoretical background;  
Section III explains the data processing and the architecture of the hybrid model;  
Section IV compares the simulation results of classical methods with those of the proposed method;  
Finally, Section V provides the conclusion.

\section{Theoretical Background}

\subsection{\textbf{\textit{Decomposition Theorem for High-Order RC Network Transfer Functions}}}

In modern integrated circuit interconnects, the parasitic resistances $R_i$ and capacitances $C_i$ distributed along metal wires form a linear time-invariant (LTI) RC network whose dynamics govern signal propagation delay, attenuation, and distortion.  To characterize this behavior, we use the Laplace-domain transfer function $H(s)$, mapping input voltage or current to output voltage or current. \cite{odabasioglu2003prima}\cite{feldmann2002efficient}\cite{antoulas2005approximation} For an arbitrary LTI RC network, the transfer function has the following essential properties:

\subsubsection{Network Topology and Mathematical Representation}
Assume the network contains $m$ independent energy-storage nodes (capacitors), leading to $m$th-order differential equations.  Applying the Laplace transform yields:
\begin{equation}
H(s)=\frac{N(s)}{D(s)}=\frac{b_0 + b_1 s + \cdots + b_{m-1}s^{m-1}}{a_0 + a_1 s + \cdots + a_m s^m}.
\end{equation}
The roots of the denominator $D(s)$ are the system poles $p_i$, satisfying:
\begin{equation}
D(s)=\prod_{i=1}^m (s - p_i), \quad p_i=-\frac{1}{\tau_i}\in\mathbb{R}^-,
\end{equation}
where $\tau_i=R_iC_i$.  The strictly negative real poles follow from the passivity of RC networks.

\subsubsection{Existence and Uniqueness of Partial Fraction Expansion}
By the fundamental theorem of algebra and the Heine–Borel theorem, any strictly proper rational function with distinct poles admits a unique partial fraction expansion\cite{ogata2009modern}:
\begin{equation}
H(s)=\sum_{i=1}^m \frac{r_i}{s - p_i},\quad r_i=\left.\frac{N(s)}{D'(s)}\right|_{s=p_i}.
\end{equation}
This result can be proven rigorously by two methods:
\begin{enumerate}
  \item \textbf{Method of Undetermined Coefficients}: Multiply both sides by $D(s)$ and equate coefficients of like powers of $s$ to obtain a linear system. The resulting Vandermonde matrix is nonsingular when poles are distinct, guaranteeing a unique solution.
  \item \textbf{Residue Theorem}: Integrating $H(s)$ around a contour enclosing all poles and applying Cauchy's residue theorem yields
  \begin{equation*}
  \oint_C H(s)\,ds = 2\pi j \sum_{i=1}^m \operatorname{Res}(H,p_i) = 2\pi j \sum_{i=1}^m r_i,
  \end{equation*}
  and analytic continuation guarantees uniqueness.
\end{enumerate}

\subsubsection{Physical Realizability Conditions}
Passivity of RC networks imposes additional constraints\cite{foster1924reactance}:
\begin{itemize}
  \item \textbf{Pole Distribution}: All poles lie strictly on the negative real axis, reflecting an overdamped response.
  \item \textbf{Residue Sign}: $r_i\in\mathbb{R}^+$, ensuring positive decaying modes.
  \item \textbf{Impulse Response}: $h(t)=\sum_{i=1}^m r_i e^{p_i t} u(t)$, illustrating multi–time-scale coupling of signal propagation.
\end{itemize}

\subsubsection{Extension to Repeated Poles}
If pole $p_i$ has multiplicity $k_i>1$ (with $\sum k_i=m$), the expansion generalizes to:
\begin{equation}
H(s)=\sum_{i=1}^q \sum_{j=1}^{k_i} \frac{r_{ij}}{(s - p_i)^j},
\end{equation}
where coefficients are given by higher-order residues:
\begin{equation}
r_{ij}=\frac{1}{(k_i-j)!}\left.\frac{d^{k_i-j}}{ds^{k_i-j}}\bigl[(s - p_i)^{k_i}H(s)\bigr]\right|_{s=p_i}.
\end{equation}

\subsubsection{Negligible Effect of Repeated Poles}
Although repeated poles introduce polynomial factors $t^{j-1}e^{p_i t}$ in the time-domain response, in practical on-chip RC networks manufacturing tolerances and layout variations make exact repeated roots unlikely. Even near-coincident poles yield higher-order residues $r_{ij}$ that are typically small, and their $t^{j-1}e^{p_i t}$ contributions decay rapidly for $p_i<0$. Hence, repeated-pole effects can be safely neglected in most modeling tasks, and single-pole expansions suffice for high-fidelity simulations.\cite{sheehan2007realizable}

This comprehensive decomposition provides a unified framework for both time-domain and frequency-domain analysis of arbitrary high-order RC interconnect networks.

\section{Data Acquisition and Preprocessing}

\subsection{Data Source and Simulation Configuration}

\subsubsection{Simulation Environment Setup}
Simulation Environment Setup
The standard cell circuit simulation environment was constructed using HSPICE, incorporating the PDK model of a 40nm CMOS technology node to ensure device parameters (e.g., threshold voltage, channel length modulation, leakage current) align with process specifications. The circuit design includes a CMOS driver unit followed by an RC signal network. The RC network parameters were generated using the SPEF (Standard Parasitic Exchange Format) netlist generator from the open-source DCTK (Digital Circuit Toolkit) platform (https://github.com/geochrist/dctk/), which ensures compliance with IEEE 1481-2009 standards for parasitic parameter extraction. This tool parses post-layout physical design information to generate RC networks that accurately reflect the interconnect characteristics of the target technology node.
\subsubsection{Stimulation Configuration}
Stimulation Configuration
The input voltage waveform was configured as an ideal step signal (0 to VDD transition). Transient simulations covered both the signal rise/fall phases (0–10 ns) and steady-state behavior.

\begin{figure}[htbp]
\centering
\begin{verbatim}
*D_NET net_0 106.518
*CONN
*I I1:Y O
*I I2:A I
*CAP
1 net_0:0 21.3035
2 net_0:1 21.3035
3 net_0:2 21.3035
4 net_0:3 21.3035
5 net_0:4 21.3035
*RES
1 I1:Y net_0:0 145.5
2 net_0:0 net_0:1 5.32588
3 net_0:1 net_0:2 5.32588
4 net_0:2 net_0:3 5.32588
5 net_0:3 net_0:4 5.32588
6 net_0:4 I2:A 145.5
*END
\end{verbatim}
\caption{An example of RC parasitic parameters in SPEF format:}
\label{fig:spef_rc}
\end{figure}

\subsection{Data Preprocessing Methods}

\subsubsection{Data Sources and Feature Definition}
The training data in this study is derived from HSPICE simulations, generating ground-truth results. The input features include voltage waveforms, device types, transient time, and RC parasitic parameters extracted from the SPEF netlist. The output target is the voltage response sequence. The feature definitions are as follows:

\textbf{Input Features:}

\begin{itemize}
    \item \textbf{Voltage Input:} The voltage time series from transient simulation, denoted as \( V(t) \in \mathbb{R}^{T} \), where the time step \( T \) is determined by the simulation time range.
    \item \textbf{Device Type:} The standard cell type, represented as categorical labels.
    \item \textbf{Transient Time:} The total simulation duration, denoted as \( t_{\text{span}} \in \mathbb{R}^{+} \).
    \item \textbf{Transfer Function Features:} Instead of directly using RC values from the SPEF netlist, we extract the transfer function from the driver to the output node and perform partial fraction decomposition. The resulting input features are a collection of triplets \( (p_i, k, A_{ik}) \in \mathbb{R} \times \mathbb{N}^+ \times \mathbb{R} \), each representing a real pole \( p_i \), its multiplicity \( k \), and the corresponding residue \( A_{ik} \). A fixed number of such terms are selected and sorted by magnitude to form a structured feature vector.
\end{itemize}

\textbf{Output Target:}
\begin{itemize}
    \item \textbf{Current Response Sequence:} The current waveform corresponding to the voltage excitation, denoted as \( I(t) \in \mathbb{R}^{T} \).
\end{itemize}

\subsubsection{Feature Normalization}
To ensure consistency across different data modalities and improve the performance of subsequent analyses, we apply specific normalization techniques tailored to each feature type:

\begin{itemize}
    \item \textbf{Voltage Waveform:} Dynamic range normalization is employed to scale the voltage values between 0 and 1, preserving the relative shape of the waveform:
    \begin{equation}
        V'(t) = \frac{V(t) - \min(V)}{\max(V) - \min(V)}
    \end{equation}
    Here, \( V(t) \) represents the original voltage at time \( t \), and \( V'(t) \) is the normalized voltage. This transformation not only ensures that all voltage inputs are on a comparable scale but also mitigates the influence of extreme values, thereby facilitating more robust learning.

    \item \textbf{Transient Time:} Given the wide range of transient times, a logarithmic compression is first applied to reduce skewness, followed by Z-score normalization:
    \begin{equation}
        t'_{\text{span}} = \frac{\log_{10}(t_{\text{span}}) - \mu_t}{\sigma_t}
    \end{equation}
    
    Here, \( t_{\text{span}} \) denotes the original transient time, and \( t'_{\text{span}} \) is the normalized value. \( \mu_t \) and \( \sigma_t \) represent the mean and standard deviation of the log-transformed transient times. This two-step process helps in mitigating the effect of outliers and ensures that the transient time feature is well-scaled for the learning process.
\end{itemize}

Overall, these normalization steps are essential for harmonizing heterogeneous data, ensuring that each feature contributes appropriately to the model. This promotes faster convergence during training and enhances the robustness of the predictive model.

\subsection{Transfer Function-Based RC Network Representation}

To more compactly and effectively capture the signal behavior of a standard-cell-driven RC interconnect, we propose a transfer function decomposition-based modeling framework. Rather than modeling internal nodes or edges, we directly extract the system-level input-output behavior of the RC network by analyzing its Laplace-domain transfer function.

\paragraph{1. Transfer Function Decomposition}

For each RC network, the voltage transfer function from the driver input to the observed output node can be expressed in its partial fraction expansion form:

\[
H(s) = \sum_{i=1}^{r} \sum_{k=1}^{m_i} \frac{A_{ik}}{(s - p_i)^k}
\]

where:
\begin{itemize}
    \item \( p_i \in \mathbb{R}^{-} \) is the \( i \)-th real pole (RC networks contain only real negative poles);
    \item \( m_i \in \mathbb{N}^{+} \) is the multiplicity (order) of pole \( p_i \);
    \item \( A_{ik} \in \mathbb{R} \) is the residue (numerator coefficient) of the \( k \)-th order term for pole \( p_i \);
    \item \( r \) is the number of distinct poles, and \( \sum_{i=1}^r m_i = n \), the total system order.
\end{itemize}

This form naturally arises from solving the nodal equations of the RC circuit in the Laplace domain and captures all time constants governing the exponential decay of the transient response.

\paragraph{2. Feature Construction: Pole-Residue Triplets}

To enable learning and generalization across circuits of different sizes, we construct a compact, structured feature vector using pole-residue triplets. Each term \( \frac{A_{ik}}{(s - p_i)^k} \) is encoded as:

\[
(p_i, k, A_{ik}) \in \mathbb{R} \times \mathbb{N}^+ \times \mathbb{R}
\]

This encoding reflects both the location and order of time constants and the strength of their contribution to the transient behavior. For a fixed representation size, the most dominant terms (by \( |A_{ik}| \)) are selected, and the rest are zero-padded.

\paragraph{3. Normalization and Sorting Strategy}

To ensure numerical stability and consistency across training samples, the following normalization strategy is adopted:
\begin{itemize}
    \item All pole locations \( p_i \) and coefficients \( A_{ik} \) are normalized with respect to the largest-magnitude pole \( |p_1| \);
    \item Pole-residue terms are sorted by \( |A_{ik}| \) to prioritize dominant dynamics;
    \item For circuits with fewer than the maximum number of terms, padding is applied using zeros or a neutral placeholder.
\end{itemize}

This representation ensures a fixed-length, physically meaningful input encoding suitable for neural network-based learning tasks, preserving the key dynamic modes of the circuit without needing explicit graph structures.

\paragraph{4. Feature Encoding with Repeated Poles}

The partial fraction expansion of a transfer function may contain repeated poles. To faithfully encode such components, each term of the form \( \frac{A_{ik}}{(s - p_i)^k} \) is represented as a triplet:

\[
(p_i, k, A_{ik}) \in \mathbb{R} \times \mathbb{N}^+ \times \mathbb{R}
\]

where:
\begin{itemize}
    \item \( p_i \): The real-valued pole location;
    \item \( k \): The multiplicity (order) of the pole;
    \item \( A_{ik} \): The numerator coefficient corresponding to the \( k \)-th order term.
\end{itemize}

This representation allows the model to explicitly capture the contribution of higher-order modes (e.g., \( t e^{-p t}, t^2 e^{-p t} \)) that arise from repeated poles in the RC transfer function. It also maintains a structured and interpretable format aligned with classical control and circuit theory.

This transfer-function-based modeling scheme transforms the RC network into a system-theoretic representation, providing a powerful and compact alternative to spatially distributed node encodings. It is particularly suitable for waveform prediction tasks where input-output response is governed by the pole-zero distribution.

\section{Hybrid Model Architecture Design}
\subsection{Overall Architecture Overview}

  - Input: Preprocessed feature vector \( X \in \mathbb{R}^{d_{in}} \) (dimension \( d_{in} = N_{time} + N_{RC} + N_{device} \))  
  
  - Output: Predicted voltage waveform sequence \( Y \in \mathbb{R}^{T} \) (time steps \( T \) determined by transient time)

This model adopts a hybrid architecture. After being uniformly encoded by the embedding layer, the input is fed in parallel into a CNN branch  and a three-layer Transformer encoder (multi-head attention to model global voltage temporal dependencies). The fused features obtained through concatenation are then passed into a three-layer Transformer decoder (masked self-attention and cross-modal attention). Finally, the predicted voltage waveform is generated through a fully connected layer.

\subsection{Input Embedding Layer}
\textbf{Function:} Map heterogeneous features to a unified latent space.

\begin{itemize}
    \item \textbf{Fully Connected (FC) Projection:}\cite{bengio2003neural}
    \begin{equation}
        H_0 = \text{ReLU}(W_e X + b_e), \quad W_e \in \mathbb{R}^{d_{embed} \times d_{in}}
    \end{equation}

    \item \textbf{Output Dimension:}
    \begin{equation}
        H_0 \in \mathbb{R}^{L \times d_{embed}}
    \end{equation}
    (L represents the sequence length, extended by time steps.)
\end{itemize}

\subsection{CNN Architecture Design}

\subsubsection{ Input Feature Adaptation}
The input to the CNN branch is the time-series feature tensor output from the embedding layer 
$\mathbf{X}_{\text{embed}} \in \mathbb{R}^{B \times L \times D}$, where $B$ is the batch size, $L$ is the sequence length, and $D$ is the embedding dimension. To adapt to the 1D convolution operation, the input tensor is dimensionally rearranged as follows:

\begin{equation}
    \mathbf{X}_{\text{reshape}} = \text{Permute}(\mathbf{X}_{\text{embed}}) \in \mathbb{R}^{B \times D \times L}
\end{equation}

This operation transforms the embedding dimension $D$ into the channel dimension, matching the convolutional layer input format $(B, C, L)$.

\subsubsection{ Convolutional Feature Extraction}\cite{bai2018empirical}
A single-layer 1D convolution kernel is employed to capture local temporal patterns, mathematically expressed as:

\begin{equation}
    \mathbf{H}_{\text{conv}} = \text{ReLU}(\mathbf{X}_{\text{reshape}} \ast \mathbf{W}_{\text{conv}} + \mathbf{b}_{\text{conv}}) \in \mathbb{R}^{B \times K \times L'}
\end{equation}

Here, $\ast$ denotes the convolution operation, $\mathbf{W}_{\text{conv}} \in \mathbb{R}^{K \times D \times S}$ represents the learnable weights ($K$ is the number of output channels, $S$ is the kernel size), and:

\begin{equation}
    L' = \left\lfloor \frac{(L - S)}{1} + 1 \right\rfloor
\end{equation}

is the post-convolution sequence length.

\subsubsection{ Feature Space Mapping}
Feature compression and dimensionality restoration are achieved through two fully connected layers:

\begin{align}
    \mathbf{H}_{\text{dense1}} &= \text{ReLU}(\mathbf{W}_1 \mathbf{H}_{\text{conv}} + \mathbf{b}_1) \in \mathbb{R}^{B \times K \times M} \\
    \mathbf{H}_{\text{out}} &= \mathbf{W}_2 \mathbf{H}_{\text{dense1}} + \mathbf{b}_2 \in \mathbb{R}^{B \times D \times L'}
\end{align}

Here, $M$ is the intermediate hidden layer dimension, $\mathbf{W}_1 \in \mathbb{R}^{M \times K}$ and $\mathbf{W}_2 \in \mathbb{R}^{D \times M}$ are learnable weight matrices. The final output is dimensionally restored to align with the Transformer branch:

\begin{equation}
    \mathbf{X}_{\text{cnn}} = \text{Permute}(\mathbf{H}_{\text{out}}) \in \mathbb{R}^{B \times L' \times D}
\end{equation}

The CNN module captures local patterns in the input sequence through convolution operations and transforms the feature space via fully connected layers, thereby capturing the correlations of the local RC network topology and enhancing the recognition of RC network information. Specifically, the local receptive field is constrained by using a convolution kernel of size \( S = 3 \)\cite{lecun1998gradient}, which limits the extraction range and forces the model to focus on the RC topological relationships of neighboring time steps. Feature interaction is lightweight as the dimensionality reduction is performed using fully connected layers rather than pooling, preserving the complete temporal resolution. Additionally, the final output dimension \( D \) is maintained equal to the input embedding dimension, ensuring compatibility for multimodal feature fusion.

\subsection{Encoder Architecture and Feature Fusion Strategy}

The Transformer encoder layer comprises a \textbf{multi-head self-attention mechanism} and a \textbf{feedforward neural network}\cite{vaswani2017attention}, both integrated with residual connections and layer normalization to facilitate deep feature learning. The mathematical formulation is as follows:

\subsubsection{Input Feature Preprocessing}
Let the input tensor be defined as $X \in \mathbb{R}^{B \times L \times D}$, where $B$ denotes the batch size, $L$ represents the sequence length, and $D$ corresponds to the embedding dimension. The encoder directly receives inputs either from the embedding layer or from the output of a preceding encoder layer.

\subsubsection{Multi-Head Self-Attention Mechanism}
Self-attention is first computed by transforming the input through learnable parameter matrices $W_Q, W_K, W_V \in \mathbb{R}^{D \times D}$ to generate the Query ($Q$), Key ($K$), and Value ($V$) matrices\cite{vaswani2017attention}:
\begin{equation}
Q = X W_Q, \quad K = X W_K, \quad V = X W_V
\end{equation}

The input is then split into $h$ attention heads, and each head computes scaled dot-product attention independently:
\begin{equation}
\text{head}_i = \text{Softmax} \left( \frac{Q_i K_i^T}{\sqrt{D/h}} \right) V_i
\end{equation}

The outputs from all attention heads are concatenated and projected as follows:
\begin{equation}
\text{MHA}(X) = \text{Concat}(\text{head}_1, \dots, \text{head}_h) W_O
\end{equation}
where $W_O \in \mathbb{R}^{D \times D}$ is the output projection matrix.

\subsubsection{Residual Connection and Layer Normalization}
To ensure stable gradient propagation and enhance training stability, a residual connection is applied, followed by layer normalization\cite{ba2016layer}\cite{srivastava2014dropout}:
\begin{equation}
X' = \text{LayerNorm}(X + \text{Dropout}(\text{MHA}(X)))
\end{equation}
This serves two key purposes:
\begin{itemize}
    \item \textbf{Residual Connection:} Helps mitigate gradient vanishing while preserving original feature representations\cite{he2016deep}.
    \item \textbf{Layer Normalization:} Improves training stability and accelerates convergence by normalizing activations across feature dimensions.
\end{itemize}

\subsubsection{Feedforward Neural Network}
A position-wise feedforward network further enhances the model's expressive power by applying two linear transformations interleaved with a non-linear activation function:
\begin{equation}
F = W_2 \cdot \text{ReLU}(W_1 X' + b_1) + b_2
\end{equation}
where $W_1 \in \mathbb{R}^{D \times d_{ff}}, W_2 \in \mathbb{R}^{d_{ff} \times D}$, and $d_{ff}$ denotes the dimensionality of the hidden layer.

\subsubsection{Second Residual Connection and Normalization}
A second residual connection and layer normalization are then applied to refine the learned representations\cite{xiong2020layer}:
\begin{equation}
X'' = \text{LayerNorm}(X' + \text{Dropout}(F))
\end{equation}

\subsubsection{Encoder Output}
The final output tensor $X'' \in \mathbb{R}^{B \times L \times D}$ is passed either to the subsequent encoder layer or to the decoder for further processing.

\subsubsection{Feature Fusion via Dot Product}
To effectively integrate global representations from the Transformer encoder and local features extracted by the CNN, we perform an element-wise dot product operation\cite{vaswani2017attention}:
\begin{equation}
    H_{\text{fusion}} = X'' \odot X_{\text{cnn}}
\end{equation}
where $X'' \in \mathbb{R}^{B \times L \times D}$ is the output from the Transformer encoder, $X_{\text{cnn}} \in \mathbb{R}^{B \times L \times D}$ is the output from the CNN branch, and $\odot$ denotes element-wise multiplication. This fusion mechanism enhances the interaction between sequential and spatial dependencies.

\subsubsection{Decoder Input Projection}
To ensure compatibility with the decoder input format, we apply a linear projection:
\begin{equation}
    H_{\text{decoder}} = W_f H_{\text{fusion}} + b_f, \quad H_{\text{decoder}} \in \mathbb{R}^{B \times L \times D}
\end{equation}
where $W_f \in \mathbb{R}^{D \times D}$ and $b_f \in \mathbb{R}^{D}$ are trainable parameters. The transformed feature representation $H_{\text{decoder}}$ is then fed into the Transformer decoder for further processing.

\subsection{Transformer Decoder Architecture}  

The Transformer decoder is responsible for generating the target sequence by leveraging masked self-attention and encoder-decoder cross-attention mechanisms. The mathematical formulation is outlined as follows.  

\subsubsection{Input Definition}  
Let the decoder input be defined as  
\begin{equation}
Y \in \mathbb{R}^{B \times T \times D},
\end{equation}
which represents the target sequence embeddings, and let the encoder output be  
\begin{equation}
H_{\text{enc}} \in \mathbb{R}^{B \times L \times D},
\end{equation}
where \(T\) denotes the target sequence length, \(L\) represents the source sequence length, and \(B\) and \(D\) indicate the batch size and embedding dimension, respectively.  

\subsubsection{Masked Multi-Head Self-Attention} \cite{vaswani2017attention}  

\textbf{Causal Masking:}  
To prevent information leakage from future time steps, a lower triangular mask is applied to the self-attention score matrix:  
\begin{equation}
\text{Mask}_{ij} =
\begin{cases}
0, & i \geq j, \\
-\infty, & i < j.
\end{cases}
\end{equation}  

\textbf{Attention Calculation:}  
The masked self-attention mechanism is computed as follows:  
\begin{equation}
    \text{Attn}_{\text{self}} = \text{Softmax}\Bigl(\frac{QK^T}{\sqrt{D}} + \text{Mask}\Bigr)V,
\end{equation}
where  
\begin{equation}
Q = K = V = Y W_{QKV},
\end{equation}
with \(W_{QKV} \in \mathbb{R}^{D \times D}\) being a learnable projection matrix. This mechanism ensures that each position in the target sequence can only attend to previous positions, preserving the autoregressive nature of decoding.  

\subsubsection{Encoder-Decoder Cross-Attention}  

\textbf{Cross-Modal Interaction:}  
The encoder-decoder cross-attention mechanism enables the decoder to incorporate contextual information from the encoder output. Specifically, the intermediate decoder representation is used as the query, while the encoder output serves as the key-value pair:  
\begin{equation}
    \text{Attn}_{\text{cross}} = \text{Softmax}\Bigl(\frac{Q_{\text{dec}}K_{\text{enc}}^T}{\sqrt{D}}\Bigr)V_{\text{enc}},
\end{equation}
where  
\begin{equation}
Q_{\text{dec}} = Y' W_Q,
\end{equation}
with \(Y'\) representing the processed decoder input and \(W_Q \in \mathbb{R}^{D \times D}\) being a trainable projection matrix. This mechanism aligns the target sequence generation with the source sequence representations, facilitating effective sequence-to-sequence modeling.  

\subsubsection{Feedforward Neural Network}  

To enhance feature representations, a two-layer feedforward network is applied:  
\begin{equation}
    F = W_2 \cdot \text{GELU}\Bigl(W_1 Y'' + b_1\Bigr) + b_2,
\end{equation}
where \(W_1 \in \mathbb{R}^{D \times d_{ff}}\), \(W_2 \in \mathbb{R}^{d_{ff} \times D}\), \(b_1 \in \mathbb{R}^{d_{ff}}\), and \(b_2 \in \mathbb{R}^{D}\). Here, \(d_{ff}\) denotes the hidden layer dimension of the feedforward network. The GELU activation function is employed to introduce non-linearity and improve gradient flow during training \cite{hendrycks2016gelu}.  

\subsubsection{Summary}  

In summary, the decoder architecture sequentially applies masked self-attention to model intra-target dependencies, encoder-decoder cross-attention to incorporate contextual representations from the encoder, and a feedforward network to refine the learned representations. This hierarchical structure ensures effective target sequence generation while preserving temporal causality.

\section{Results}
\subsection{Scalability Across Network Sizes}

To evaluate the scalability of our neural network model with respect to increasing circuit complexity, we constructed a set of networks with varying numbers of 
RC nets, ranging from 2 to 100. For each network configuration, we simulated the voltage response using both our neural network predictor and a SPICE-based transient solver.

The results, shown in Table~\ref{tab:n_vs_rmse}, indicate that the RMSE generally decreases as $n$ increases from 2 to 80, suggesting that the ensemble of more sub-networks enhances the model’s representation capability. The best performance was achieved at $n = 80$ with an RMSE of \textbf{0.0020}, followed closely by $n = 70$ and $n = 90$. However, when $n$ increased to 100, the RMSE sharply rose to \textbf{0.0098}, indicating possible overfitting or optimization instability. These findings demonstrate that while increasing $n$ generally improves performance, there exists an optimal range ($n \in [70,90]$) beyond which the model may degrade.

\begin{table}[ht]
\centering
\caption{Validation RMSE for different numbers of subnetworks $n$ on the RC\_1 dataset.}
\label{tab:n_vs_rmse}
\begin{tabular}{c|c}
\hline
\textbf{Number of Networks ($n$)} & \textbf{Validation RMSE} \\
\hline
2   & 0.0079 \\
5   & 0.0043 \\
10  & 0.0074 \\
15  & 0.0045 \\
20  & 0.0029 \\
30  & 0.0042 \\
40  & 0.0057 \\
50  & 0.0032 \\
60  & 0.0067 \\
70  & 0.0021 \\
80  & \textbf{0.0020} \\
90  & 0.0023 \\
100 & \textbf{0.0098} \\
\hline
\end{tabular}
\end{table}

Figure below shows the root mean squared error (RMSE) between the predicted and ground-truth voltage waveforms as a function of the network size. Notably, the prediction accuracy remains consistent even as the number of internal nodes increases, with only a modest rise in error for the largest networks. This demonstrates that the model can generalize well to large-scale RC networks, indicating its potential for hierarchical or full-chip level timing analysis.

\begin{figure}[h]
    \centering
    \includegraphics[width=1\linewidth]{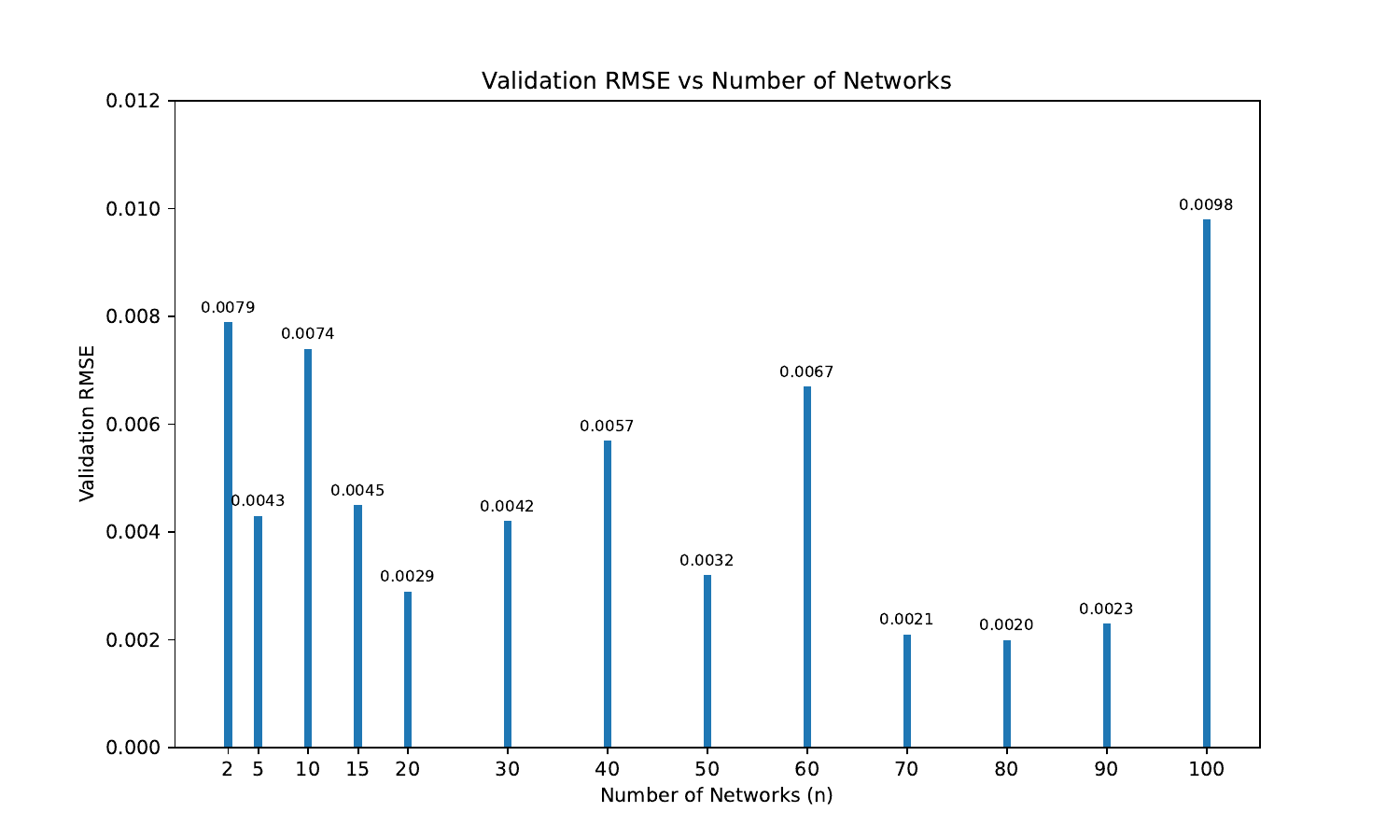}
    \caption{rmse\_vs\_n}
    \label{fig:enter-label}
\end{figure}

\subsection{Accuracy of Voltage Response Prediction}
We further quantified the accuracy of our model by comparing its output against SPICE simulations across a wide range of RC networks with diverse topologies and parameter settings. The comparison was performed on a held-out test set of 100 RC configurations not seen during training.

\begin{figure}[ht]
    \centering
    \includegraphics[width=0.48\textwidth]{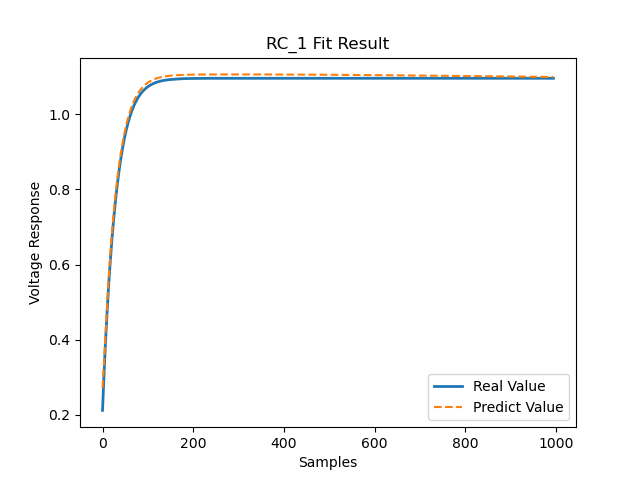}
    \caption{Comparison of voltage response between SPICE simulation and neural network prediction on a representative RC network.}
    \label{fig:rc1_fit_result}
\end{figure}

Figure~\ref{fig:rc1_fit_result} shows an example voltage trace comparison between the SPICE ground truth and the neural network prediction for a representative circuit. The predicted waveform (orange dashed line) closely matches the true response (blue solid line), accurately capturing the rise behavior, steady-state voltage level, and response curvature. Notably, the model predicts the early-stage transient response with high fidelity, which is critical in timing-sensitive applications.

Across the full test set, the RMSE was measured at 0.009813355. These results confirm that the proposed model achieves high-fidelity predictions, rivaling traditional circuit simulators in accuracy while significantly reducing inference time.

\subsection{Training time and Infer time Comparison}

As shown in Table~\ref{tab:train_time_vs_n}, the training time remains relatively stable across different numbers of sub-networks $n$, ranging from approximately 77 to 80 seconds. This indicates that the increase in model capacity due to higher $n$ values does not significantly impact the training runtime, likely due to efficient batching and parallelization during training.

\begin{table}[ht]
\centering
\caption{Training time (in seconds) for different number of sub-networks $n$.}
\label{tab:train_time_vs_n}
\begin{tabular}{c|c}
\hline
\textbf{Number of Networks ($n$)} & \textbf{Training Time (s)} \\
\hline
2   & 78.19 \\
5   & 79.25 \\
10  & 77.53 \\
15  & 79.39 \\
20  & 77.59 \\
30  & 77.69 \\
40  & 77.93 \\
50  & 78.24 \\
60  & 78.89 \\
70  & 78.44 \\
80  & 78.64 \\
90  & 80.43 \\
100 & 79.54 \\
\hline
\end{tabular}
\end{table}

\begin{figure}[h]
    \centering
    \includegraphics[width=1\linewidth]{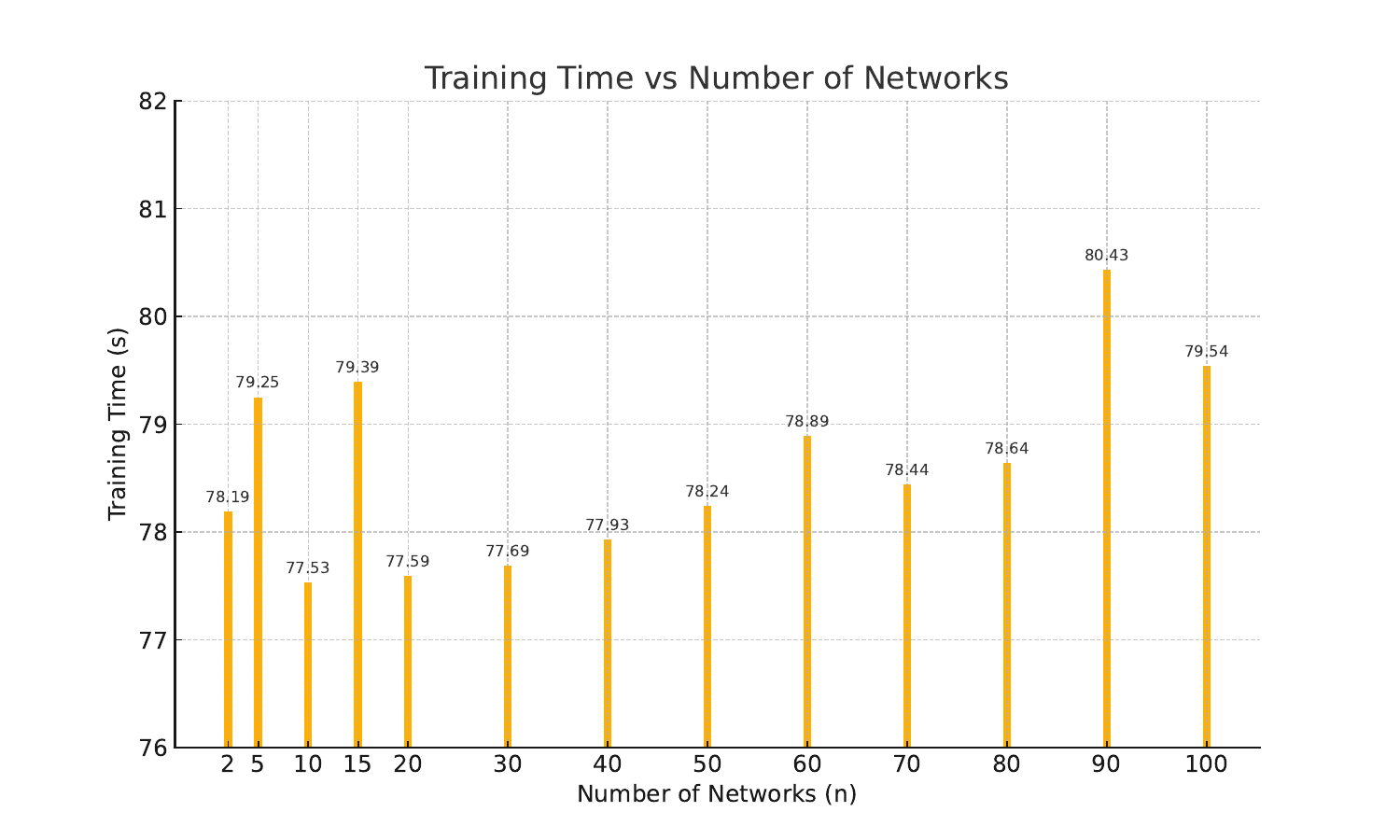}
    \caption{Total\_Time\_vs\_n}
    \label{fig:enter-label}
\end{figure}

\section{Future Challenges and Directions for our model}
\section{Future Challenges and Directions}

Despite the promising results achieved by our neural predictor, several challenges remain before it can be adopted in broader circuit analysis workflows. In this section, we discuss key limitations of the current approach and highlight potential directions for future research.

\subsection{Limited Generalization Across Topologies}

One of the main limitations of the current model is its restricted ability to generalize across circuits with different topological complexities. The model is trained on RC networks with fixed-size inputs and relatively homogeneous structure, which limits its applicability to networks with varied node counts, branch configurations, or dynamic component distributions. When applied to a topology unseen during training, the prediction accuracy tends to degrade significantly. Addressing this challenge requires the development of topology-aware or topology-invariant architectures. For instance, graph-based neural networks (GNNs) that operate directly on circuit graphs could offer a more flexible representation, enabling better generalization across a wider variety of network configurations.

\subsection{Quadratic Complexity of Transformer Inference}

The current architecture relies on the standard Transformer encoder-decoder framework, which incurs quadratic computational complexity $\mathcal{O}(n^2)$ with respect to the input sequence length $n$ due to the self-attention mechanism. Although this cost is manageable for small circuits, it becomes prohibitive for larger-scale networks with hundreds or thousands of nodes. Reducing this bottleneck is essential for scaling to industrial-grade circuit sizes. Recent advances in efficient Transformers, such as Linformer, Performer, or Longformer, which reduce attention complexity to linear or sub-quadratic time, present viable alternatives that could be integrated into future versions of the model.

\subsection{Temporal Alignment and Stability Under Noise}

Another challenge lies in ensuring the temporal alignment of predicted and true voltage waveforms, especially in high-frequency transients. While the model performs well under nominal conditions, it may exhibit phase drift or temporal misalignment when circuit parameters vary significantly or when input signals are noisy. This suggests the need for explicit mechanisms to model temporal causality and uncertainty. Incorporating recurrent components (e.g., gated RNNs) or probabilistic modeling (e.g., Bayesian neural nets) could enhance the model's robustness to such variations.

\subsection{Lack of Physical Interpretability and Control}

Finally, like many black-box neural models, the current predictor lacks interpretability and offers limited control over specific electrical behaviors (e.g., delay tuning, overshoot suppression). This poses challenges for deployment in design-in-the-loop settings where precise control is required. Future work may consider embedding circuit-theoretic constraints or physical priors directly into the model architecture or training loss. Hybrid models that combine learnable components with analytical solvers could strike a balance between accuracy, efficiency, and interpretability.

\vspace{0.5em}

In summary, while our work demonstrates that neural networks can serve as fast and accurate circuit response predictors, substantial efforts are still needed to bridge the gap toward general-purpose, scalable, and controllable neural circuit solvers. We leave these directions for future exploration.

\section{Conclusion}

In this work, we presented an initial demonstration of a neural network-based surrogate model for fast prediction of voltage responses in RC networks. The proposed architecture integrates Transformer encoders and convolutional modules to simultaneously capture temporal dependencies and spatial features across network nodes. By leveraging structured input representations of circuit topology and component values, the model is able to learn complex transient behaviors with high accuracy.

Through experiments on synthetic RC networks with varying parameters, we showed that the model can closely approximate the voltage trajectories obtained from SPICE simulations, while significantly reducing inference time. Our results confirm the feasibility of applying deep learning to analog circuit modeling tasks, and highlight the potential of hybrid architectures in capturing both global context and local signal dynamics.

This first-stage demonstration provides a foundation for future extensions toward more generalized, scalable, and interpretable neural solvers. In subsequent work, we aim to further enhance the model's adaptability to diverse topologies and explore integration with traditional simulation frameworks.


%

\ifCLASSOPTIONcaptionsoff
  \newpage
\fi



%
\bibliographystyle{ieeetr}
\bibliography{main.bbl}

\end{document}